\documentclass[pra,aps,twocolumn,footinbib,floatfix,showpacs,superscriptaddress]{revtex4}%
\usepackage{times,amsmath}
\usepackage{epsfig}
\usepackage{amsmath}
\usepackage{amsfonts}
\usepackage{amssymb}
\usepackage{graphicx}%
\providecommand{\U}[1]{\protect\rule{.1in}{.1in}}
\newcommand{\TMYAG}{Tm$^{3+}$:YAG }
\begin{document}
\title{Why the two-pulse photon echo is not a good quantum memory protocol}
\author{J\'er\^ome Ruggiero}
\affiliation{Laboratoire Aim\'e Cotton, CNRS-UPR 3321, Univ. Paris-Sud, B\^at. 505, 91405 Orsay cedex, France}
\author{Jean-Louis Le~Gou\"et}
\affiliation{Laboratoire Aim\'e Cotton, CNRS-UPR 3321, Univ. Paris-Sud, B\^at. 505, 91405 Orsay cedex, France}
\author{Christoph Simon}
\affiliation{Group of Applied Physics, University of Geneva, CH-1211 Geneva 4, Switzerland}
\author{Thierry Chaneli\`ere}
\email{thierry.chaneliere@lac.u-psud.fr}
\affiliation{Laboratoire Aim\'e Cotton, CNRS-UPR 3321, Univ. Paris-Sud, B\^at. 505, 91405 Orsay cedex, France}

\pacs{03.67.-a, 32.80.Qk, 42.50.Md, 42.50.–p, 42.50.Gy, 42.50.Md}

\begin{abstract}
We consider in this paper a two-pulse photon echo sequence as a potential quantum light storage protocol. It is widely believed that a two-pulse scheme should lead to very low efficiency and is then not relevant for this specific application. We show experimentally by using a Tm${}^{3+}$:YAG crystal that such a protocol is on contrary very efficient and even too efficient to be considered as a good quantum storage protocol. Our experimental work allows us to point out on one side the real limitations of this scheme and on the other side its benefits which can be a source of inspiration to conceive more promising procedures with rare-earth ion doped crystals.
\end{abstract}
\maketitle

\section{Introduction}

The prospect of quantum light storage in solids motivates us to reconsider the interaction of light and matter at the single quantum level. Historically coherent transient phenomena appeared very active because they are primarily based on two-level system absorption. The community naturally focused on rare-earth ion doped crystals (REIC) instead of atomic vapors because the storage time can be long as well and is not limited by the atomic diffusion whatsoever. Recent proposals are of course inspired by previous realizations of classical light memories \cite{mossbergOL83} and all-optical processing more generally \cite{babbitAO94}. In this lineage recent progresses toward quantum storage are involving common physical ingredients. Because of the large inhomogeneous broadening, any light retrieval is intimately related to a dipole rephasing. Surprisingly the very well known conventional photon echo has not been considered as a promising quantum protocol despite impressive realizations in the classical domain. A major identified drawback is the low efficiency of the process \cite{moiseevPRL01, moiseevJOB03}. This has been a commonly admitted idea since the pioneering studies of the 60's \cite{hartmann64}. Although we roughly adhere to this statement, we would like to point out that high retrieval efficiency was already observed in a specific regime of two-pulse photon echo (2PE) \cite{cornishPRA98,cornishOL00}, which might pave the way to quantum light storage. Our experimental work consists of a clear observation of the large predicted efficiency in this regime we first need to precisely defined. Our analysis is also stressing clearly what are the physical ingredients which lead to this result. Our study is definitely placed in the prospect of a quantum memory. With respect to other protocols, we finally clarify the advantages and drawbacks of this technique that should be considered as a general tool for coherent manipulations.

As mentioned before, the investigation of classical light storage largely paved the way toward their quantum equivalent. REIC have shown interesting processing capabilities especially with all-optical control \cite{babbitAO94,tian}. As derived from the conventional photon echo, these techniques are based on an optical manipulation of the coherences. Experiments largely benefit from the agility of the laser controlling the crystalline processor \cite{crozatier}. This convenience would be still appreciable for manipulation at the quantum level. The 2PE time-to-bandwidth product properties should also be emphasized. In the 2PE process, this parameter, critical for information processing applications, is not limited by the memory opacity, in contrast with the most promising quantum storage protocols involving REIC, namely the "stopped-light" approach \cite{EITlukin,turukhin2002} or the "controlled reversible inhomogeneous broadening" (CRIB) procedure \cite{crib1,alexander2006,crib3, hetet07}.     
Finally, contrary to the above storage protocols, doesn't require any initial state preparation. A spectral selection within the inhomogeneous broadening is in a sense build-in because of the selective excitation of the first incoming pulse that we define as the signal. The 2PE has the singular advantage to rephase a random distribution of level shifts without any assumption on the source of inhomogeneity. By clarifying the characteristics of the 2PE, we aspire to a deeper understanding of the atomic coherences optical manipulation.

Since the observation \cite{hartmann64} and the interpretation \cite{hartmann66} of the photon echo, the quantitative comparison with the observed efficiency has been widely studied. At the basis of data processing application, the interest has been renewed relatively recently \cite{cornishPRA98, mossbergPRA99}. A realistic approach of the problem usually requires a numerical resolution of the Bloch-Maxwell equations \cite{mossbergOC98}. This gives a solid interpretation of the experimental data \cite{mossbergPRA99}. As a matter of fact investigating the 2PE for quantum light storage is strongly simplifying the problem. As compared to the canonical 2PE where a $\pi/2$ excitation pulse is followed by a $\pi$ rephasing pulse, memory-like version of this scheme would first involve a very weak signal pulse. In that very specific case as pointed out by Tsang et al. \cite{cornishJosab03}, one can derive an analytic solution for the efficiency. In this paper, we will first deduce these equations from a simple physical interpretation of the scheme. We will clearly specify the underlying assumptions to be verified in practice. We will then compare these calculations to the experiment. A detailed analysis of our observations allows us to conclude and place this work in the context of a quantum memory by comparison to other storage protocols.

\section{Efficiency of the protocol}
This subject has been covered  by a wide range of literature \cite{cornishPRA98, cornishOL00, cornishJosab03, mossbergPRA99, mossbergOC98}. Nevertheless it is relatively easy to derive these equations based on simple physical arguments. The interaction of light pulses with our medium is well described by the Maxwell-Bloch equations assuming the slowly varying amplitude and the rotating wave approximations.

\begin{equation}\label{MB}
\begin{array}{ll}
\partial_z\Omega(z,t)&=-\displaystyle\frac{\alpha}{2\pi}\int d\omega_{ab}\mathrm{v}(\omega_{ab};z,t)\\[0.4cm]
\partial_t\mathrm{u}(\omega_{ab};z,t)&=-\Delta \mathrm{v}(\omega_{ab};z,t)\\[0.3cm]
\partial_t\mathrm{v}(\omega_{ab};z,t)&=-\Omega(z,t)\mathrm{w}(\omega_{ab};z,t)+\Delta \mathrm{u}(\omega_{ab};z,t)\\[0.3cm]
\partial_t\mathrm{w}(\omega_{ab};z,t)&=\Omega(z,t)\mathrm{v}(\omega_{ab};z,t)
\end{array}
\end{equation}
where $\Delta=\omega_{ab}-\omega_L$ is the detuning, $\Omega$ is the Rabi frequency of the field under consideration and $(\mathrm{u},\mathrm{v},\mathrm{w})$ the three components of the Bloch vector. The decay of the coherences and the population is assumed to be negligible. We have dropped the usual term $\partial_t\Omega(z,t)$ because realistically the spatial extension of the pulse is always much longer the length of our crystal.

\begin{figure}[pth]
\includegraphics[width=6cm]{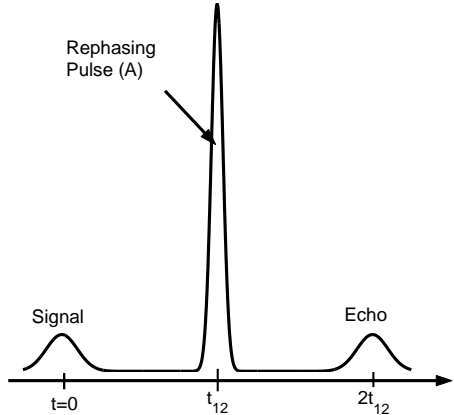}\caption{Outline of the time sequence. The signal is first absorbed in the medium. After a time $t_{12}$ a rephasing pulse of large area A induces a build-up of the coherence at the time $2t_{12}$ and give rise to the photon echo. We address the problem in the weak signal and the echo regime, which shall be satisfied in quantum field conditions.}
\label{fig:seqT}%
\end{figure}

To describe the broadest range of situation a numerical resolution of the system is usually necessary \cite{mossbergOC98}. This is not our approach. A sketch of the time sequence is depicted in Fig.\ref{fig:seqT} and looks like any 2PE sequence. Nevertheless within the prospect of quantum storage, the signal and the echo are assumed to be weak. This greatly simplifies the description \cite{crisp} essentially because these two fields do not modify the population difference $\mathrm{w}(\omega_{ab};z,t)$ that is not time dependent anymore. This is the small area approximation where the Maxwell-Bloch system can be linearized \cite{crisp}.

\subsection{Calculation in the weak signal limit}

If the three pulses are well separated in time, one can consider them independently. The signal, the rephasing pulse and the echo are respectively centered on $t=0$, $t_{12}$ and $2t_{12}$. The incoming signal and outgoing echo Rabi frequencies are respectively denoted $\mathcal{S}$ and $\mathcal{E}$. The propagation of $\mathcal{S}$ is simply described by an absorption law if the atoms are initially in the ground state $\mathrm{w}(\omega_{ab};z,-\infty)=\mathrm{w}(\omega_{ab};z,0)=-1$ \cite{crisp}:
\begin{equation}\label{AbsS}
\partial_z\mathcal{S}(z,t)=\mathrm{w}(\omega_{ab};z,0) \frac{\alpha}{2} \mathcal{S}(z,t) = -\frac{\alpha}{2} \mathcal{S}(z,t) 
\end{equation}

The weak echo is expected to behave in a similar way, except the medium has been previously excited and modified by the signal and the rehasing pulse. Therefore the echo equation reads as:
\begin{equation}\label{AbsE}
\begin{array}{ll}
\partial_z\mathcal{E}(z,t)=&\mathrm{w}(\omega_{ab};z,2t_{12}) \frac{\alpha}{2} \mathcal{E}(z,t) \\
&-\displaystyle\frac{\alpha}{2\pi}\int d\omega_{ab}\mathrm{v_E}(\omega_{ab};z,t)
\end{array}
\end{equation}
The coherence $\mathrm{v_E}$, resulting from interaction with the first two pulses, evolves freely within the time interval $t_{12} \rightarrow 2t_{12}$. The population $\mathrm{w}(\omega_{ab};z,2t_{12})$ has been affected by the rephasing pulse. We assume the rephasing pulse is much shorter than $\mathcal{S}$. Therefore the population is uniformly modified by the rephasing pulse all over the spectral interval initially excited by $\mathcal{S}$. We will see in \ref{sec:exp} how this constraint is treated experimentally. A second benefit of this assumption is that the rephasing pulse can be consider as instantaneous (time $t_{12}$) and then uniformly modifying the coherences. The physical interpretation of Eq.\ref{AbsE} is based on the generation of $\mathcal{E}$ by the macroscopic dipole that builds up when the coherences $\mathrm{v_E}$ get phased together. The growing field propagates through the medium characterized by the uniform population difference $\mathrm{w}(\omega_{ab};z,2t_{12})$. 

\begin{figure}[pth]
\includegraphics[width=8cm]{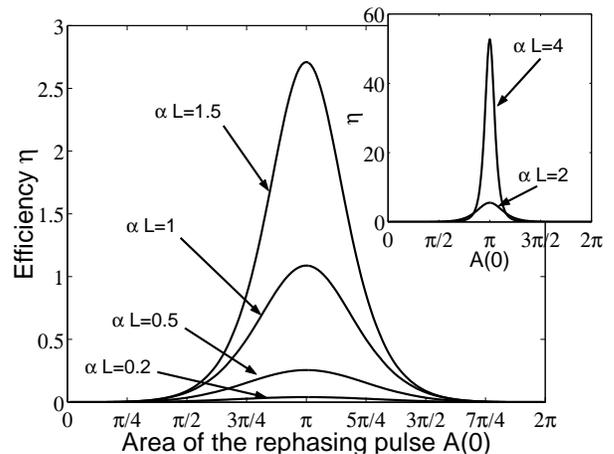}\caption{Efficiency of the 2PE as a function of the rephasing pulse area $A\left(0\right)$ for different optical thickness $\alpha L$. Inset: for larger $\alpha L$, the efficiency is much larger than unity at the maximum rephasing.}
\label{fig:PlotEtaVSAi}%
\end{figure}

We aim at reducing Eq.\ref{AbsE} by expressing the atomic quantities in terms of the optical fields only. We shall be left with an equation of propagation for $\mathcal{E}$. The echo efficiency will be deduced from the solution of this equation. We first have to track the excitation of the coherences by the signal, then their modification by the rephasing pulse and finally their free evolution toward the echo emission. The problem is addressed locally, at position $z$. The signal $\mathcal{S}$ excites the atoms that initially all sit in the ground state:
\begin{equation}
\begin{array}{ll}
\mathrm{u}(\omega_{ab};z,t)&+i\mathrm{v}(\omega_{ab};z,t)=\\[0.3cm]
&i\exp\left(i\Delta t\right) \displaystyle\int_{-\infty}^{t}\mathcal{S}(z,\tau)\exp\left(-i\Delta\tau\right)\mathrm{d}\tau
\end{array}
\end{equation}
At a certain time $t$ between $0$ and $t_{12}$ the signal field is off. We then recognize the Fourier transform of $\mathcal{S}$ written $\widetilde {\mathcal{S}}$, the $\exp\left(i\Delta t\right)$ accounts for the free evolution during this interval.
\begin{equation} \label{UV}
\mathrm{u}(\omega_{ab};z,t)+i\mathrm{v}(\omega_{ab};z,t)= i\exp\left(i\Delta t\right) \widetilde {\mathcal{S}}\left(z,\Delta \right)
\end{equation}
This expression represents the evolution of the coherence after the initial absorption process and before the rephasing pulse. Next, we calculate this strong pulse effect on the coherences (how $\mathrm{v_E}$ is related to $\mathrm{v}$) and on the population, which modifies the echo propagation (Eq.\ref{AbsE}). This can be done analytically by integrating the Bloch-Maxwell equations (see ref. \cite{cornishJosab03}). Nevertheless the results are relatively intuitive at the end and can be derived from simple physical ingredients. This can be done first of all in the specific case of a $\pi$-rephasing pulse, the more general case of an area $A(z)$ for the strong pulse can be solved by introducing "by hand" geometrical factors. All over the spectral interval excited by the signal pulse, the rephasing pulse is assumed to behave as a $\pi$-pulse. This corresponds to a brief pulse assumption. The experimental fulfillment of this condition will be addressed in \ref{sec:exp}. A $\pi$-pulse simply drives the Bloch vector by a rotation of $\pi$ around an equatorial axis. On the one hand, along the population axis it corresponds to an inversion from $-1$ to $\mathrm{w}(\omega_{ab};z,2t_{12})=1$ at the time $t_{12}$. On the other hand, it transforms the coherences $\mathrm{v}(\omega_{ab};z,t_{12})\rightarrow-\mathrm{v}(\omega_{ab};z,t_{12})$, $\mathrm{u}(\omega_{ab};z,t_{12})$ stays the same (complex conjugation of $\mathrm{u}+i\mathrm{v}$). Right after the rephasing pulse, Eq. \ref{UV} becomes
\begin{equation} \label{UEVE}
\begin{array}{ll}
\mathrm{u_E}(\omega_{ab};z,t)&+i\mathrm{v_E}(\omega_{ab};z,t)=\\[0.3cm]
& i\exp\left(i\Delta \left(t-t_{12}\right)\right) \widetilde {\mathcal{S}}^*\left(z,\Delta \right)\exp\left(-i\Delta t_{12}\right)
\end{array}
\end{equation}
where the complex conjugation sign $*$ accounts for the rephasing transformation. The coherences are freely evolving after $t_{12}$. One can now write the propagation equation of the echo (Eq.\ref{AbsE}) by including the direct influence of the signal on $\mathrm{v_E}(\omega_{ab};z,t)$. The signal field being assumed to be a real number, one recognizes the time-reversed signal $\mathcal{S}(z,2t_{12}-t)$ whose $z$ dependency is given by an absorption law $\mathcal{S}(z,2t_{12}-t)=\mathcal{S}(0,2t_{12}-t)\exp\left(-\alpha z\right)$(Eq.\ref{AbsS})
\begin{equation}\label{EfinalPI}
\partial_z\mathcal{E}(z,t)=+\frac{\alpha}{2} \mathcal{E}(z,t)-\alpha \mathcal{S}(z,2t_{12}-t) 
\end{equation}
The signal field acts as a source and generates the echo that propagates in an inverted medium. This gives the equation of propagation for a $\pi$-rephasing pulse. It is now rather easy to account for an imperfect rephasing. More generally, a $A(z)$-area strong pulse drives an $A(z)$-rotation of the Bloch vector. The population is not fully inverted anymore: $\mathrm{w}(\omega_{ab};z,2t_{12})=-\cos A\left(z\right)$. The rotation of the coherences is also incomplete and limited to $\left[1-\cos\left(A\left(z\right)\right)\right]/2$ of its maximum value. This factors are purely geometrical and are interpreted as projections on the Bloch sphere. We finally get the general analytic expression for the efficiency. This expression has been previously derived by Tsang \textit{et al.} \cite[Eq.(40)]{cornishJosab03} by integrating the Bloch-Maxwell equations. Here we simply focus on the two crucial stages, the absorption of the signal on one side and the re-emission of the echo on the other side. The rephasing pulse in between is interpreted as an instantaneous manipulation of the Bloch vector.
\begin{equation}\label{Efinal}
\begin{array}{ll}
\displaystyle\partial_z\mathcal{E}(z,t)=&-\cos\left(A\left(z\right)\right)\frac{\alpha}{2} \mathcal{E}(z,t)\\[0.4cm]
&-\displaystyle\frac{1-\cos\left(A\left(z\right)\right)}{2}\alpha \mathcal{S}(z,2t_{12}-t)
\end{array}
\end{equation}
There is an underlying assumption here: the rephasing pulse is very brief and is then fully covering the spectral range of excitation. As a consequence the echo is not deformed as compared to the signal, it is only time reversed. The $A(z)$-rotation on the Bloch sphere is also uniform and doesn't depend on $\Delta$. The $z$-dependency of the area accounts for the propagation of the strong pulse itself. This is usually a complicated problem but in that case we are only interested in the propagation of the area. The result is remarkably simple and is given by the Area Theorem of McCall \& Hahn \cite{mccall1967}. We don't have to know the exact temporal shape through the propagation because the area is the relevant quantity for the rephasing pulse and is simply given by
\begin{equation}\label{Area}
\partial_z A\left(z\right)=-\frac{\alpha}{2}\;\sin A\left(z\right)
\end{equation}
It can be solve analytically for a given $A\left(0\right)$. A straightforward integration of Eq.\ref{Efinal} allows us to calculate the retrieval efficiency $\eta$ as a function of the optical thickness $\alpha L$ where $L$ is the length of the medium:
\begin{equation}\label{eta}
\begin{array}{ll}
\displaystyle\eta\left(A\left(0\right),\alpha L\right)&=\displaystyle\left(\frac{\mathcal{E}(L,t)}{\mathcal{S}(0,2t_{12}-t)}\right)^2\\[0.4cm]
&=\displaystyle\left(\frac{2\sinh\left(\alpha L\right)}{1+\exp\left(\alpha L\right)\cot^2\left(A\left(0\right)/2\right)}\right)^2
\end{array}
\end{equation}
For different optical thickness $\alpha L$, we plot the efficiency as a function of the rephasing pulse area $A\left(0\right)$ in Fig.\ref{fig:PlotEtaVSAi}.
We observe in that case that the efficiency is strongly depending on the optical thickness. When it is low, the efficiency is weak essentially because the signal is poorly absorbed. The efficiency is then a sinus-like function and the rephasing area directly accounts for an imperfect rotation on the Bloch sphere. On the other side, at large optical thickness, the efficiency can be much larger than unity but only on a narrow window around a $\pi$. This specific situation is certainly the most interesting because the signal in fully mapped into the medium (large optical thickness) and the efficiency is exceptionally large. The $\pi$-rephasing pulse have very particular properties that can be interpreted independently to physically understand this behavior.

\subsection{Specificity of the $\pi$-rephasing pulse}
\label{specPI}
The case of an exact $\pi$-rephasing pulse is relatively straightforward. More fundamentally it allows us to understand the expected efficiency at large optical thickness. We simply derived the propagation equation (\ref{EfinalPI}) for the echo by assuming that locally, at position $z$, the area is exactly $\pi$. Now we have to examine the $\pi$-pulse propagation inside the medium. This is a very specific situation. According to Eq.\ref{Area} a $\pi$-pulse preserves its initial area throughout the medium. Even if the energy is absorbed, the area is conserved as the pulse stretches temporally~\cite{ruggiero}. Roughly speaking, if the pulse is elongated by a factor $r$, the amplitude (Rabi frequency) is reduced by $r$ to conserve the area and the energy decreases by a factor $r$. This alteration is a pure coherent propagation effect.

The propagation equation \ref{EfinalPI} is then valid at any position $z$ and easily gives the efficiency. This expression is consistent with our general formula Eq. \ref{eta}:
\begin{equation}\label{etaPi}
\eta\left(\pi,\alpha L\right)=\left[\exp\left(\alpha L/2\right)-\exp\left(-\alpha L/2\right)\right]^2
\end{equation}
At large optical thickness, the efficiency is much larger than unity and grows exponentially. This is relatively counter-intuitive. The echo efficiency is generally observed to be low, which is usually assigned to absorption. As mentioned before the $\pi$-rephasing pulse retains its area along the propagation. In other word, the medium is completely inverted: the echo is emitted in an amplifying medium. This explains why the echo is gaining exponentially. Practically, the assumptions we made, such as 1-dimension, infinite plane wave geometry, will be difficult to satisfy. Any divergence from the ideal theoretical frame shall affect the echo efficiency.   

\begin{figure}[pth]
\includegraphics[width=8cm]{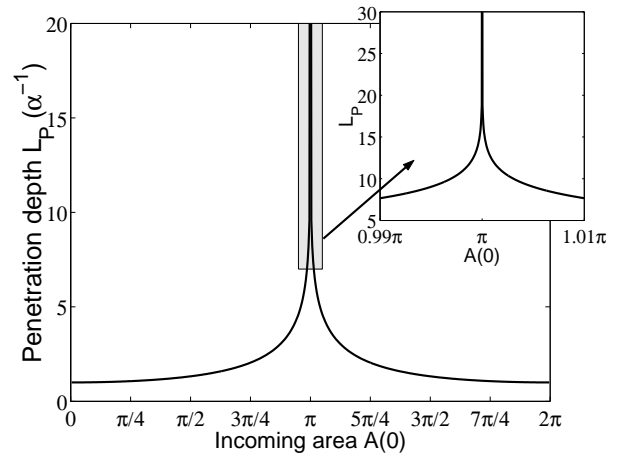}\caption{Penetration depth $L_{P}$ of a strong pulse in unit of $\alpha^{-1}$ as a function of the incoming area. Inset: Penetration of pulses with area of $\pi$ plus or minus 1\%. For small area pulses, it is simply given by an absorption law $L_{P}=\alpha^{-1}$.}
\label{fig:PlotLeff}%
\end{figure}

The $\pi$-pulse propagation is not only unusual, it is also a singular solution of the Area Theorem. The $\pi$-solution is indeed not stable because any area slightly lower (or larger) than $\pi$ will decrease (or increase resp.) toward $0$ (or $2\pi$ resp.) \cite{mccall1967,ruggiero}. Even so, the pulses with an area close to $\pi$ can propagate deeper inside the medium than a weak pulse. To see that we plot in Fig. \ref{fig:PlotLeff} the penetration depth $L_{P}$ at which the incoming area is divided by $1/\sqrt{e}$ (the curve is symmetrized around $\pi$ to account for the deviation toward $2\pi$ of pulses larger than $\pi$).

For small areas, this length is $\alpha^{-1}$ as expected \cite{crisp}. As one gets closer to $\pi$, the pulse can propagate deeper inside the medium. For example, if the incoming area is controlled at the 1\% level (inset Fig. \ref{fig:PlotLeff}), the penetration depth is larger than $7 \alpha^{-1}$. The curve is very narrow about $\pi$ which shows a high sensitivity for the propagation. Realistically a well-controlled $\pi$-pulse should induce a population inversion much deeper inside the medium than the absorption length $\alpha^{-1}$. If this depth is larger than the optical thickness of the medium, it will be fully inverted. However, the deeper the $\pi$-pulse propagates through the absorbing medium, the more it is stretched, since it has to keep a constant area while losing energy. The pulse bandwidth shrinks accordingly, making the pulse act as a $\pi$-pulse on a reduced spectral interval~\cite{ruggiero}.     

Based on this analysis, we expect to observe two remarkable qualitative features. The efficiency should be very high at large optical thickness and strongly depending on the area of the rephasing pulse.

\section{Experiments}
\label{sec:exp}
A 2PE experiment can be performed in any system were a transient phenomena can be observed. Here we use a thulium-doped yttrium aluminum garnet (YAG), cooled down to cryogenic temperature. The long optical coherence time makes it particularly attractive for quantum storage application. We will now briefly describe the experimental set-up and focus on the precautions we take to satisfy the assumptions introduced previously. 
\subsection{Experimental set-up}
Our 0.5\% \TMYAG crystal is immersed in liquid helium at 1.4K. The coherence time of the $^{3}$H$_{6}$(0) to $^{3}$H$_{4}%
$(0) transition is typically $T_2=50\mu s$ in these conditions. The crystal is oriented and cut in order to propagate along the [1$\mathrm{\bar{1}}$0] direction. Along this axis, the length is 5mm and the optical thickness $\alpha L=5$. The laser polarisation is parallel to [111] to maximize the Rabi frequency \cite{ConeNutAngle}. The laser system is operating at 793~nm, stabilized on a high-finesse Fabry-Perot cavity with the Pound-Drever-Hall technique (200Hz over 10ms) \cite{crozatier2004}. The laser is split in two independent beams. Temporal shaping is achieved by two acousto-optic modulators (AOM) controlled by a dual channel arbitrary waveform generator (Tektronix AWG520). Both beams are injected into two single mode fibers. Before recombination on a beamsplitter, we use expanders to independently manage their waists inside the crystal. After the sample, the signal is collected in a single mode fiber terminated by a photodiode.

The signal is supposed to mimic a weak quantum field, so this pulse should verify the small area approximation. The signal beam is in practice much weaker than the rephasing one. At the maximum, there is a 36dB power difference between the two. More precisely, the signal area is kept constant at 9\% of $\pi$ and we vary the rephasing pulse area $A\left(0\right)$ from 0 to $~3\pi$.

Keeping the rephasing pulse significantly briefer than the signal pulse is the most stringent condition we have to satisfy. This is required to maintain a uniform coverage of the rephasing process over the signal excitation bandwidth. To do so, we use gaussian-shaped signal (duration $\mathrm{2.1\mu s}$) and rephasing pulses. On the one hand a gaussian pulse is spectrally narrower than a rectangular pulse with the same duration. On the other hand we observed a gaussian $\pi$-pulse undergoes less temporal streching than a gaussian one after propagation through an absorbing medium. We set the rephasing pulse $\approx2.5$ times shorter than the signal. This value is slightly fluctuating depending on the rephasing amplitude value. Changing the AOM driving power marginally impacts on the pulse shape. A much shorter pulse would be preferable but we are limited by the available power (few milliwatts) to ensure a significant area in a reasonable time.

One last point we ignored so far is the transverse dimension of the beams. To be consistent with the 1-dimension theory, the power of the rephasing beam should be constant over the spatial extension of the signal. The signal beam waist ($\mathrm{17 \mu m}$) is then chosen to be 2 times smaller than the waist of the rephasing beam ($\mathrm{35 \mu m}$). This is the same overlap argument we used in the spectral domain. 

\subsection{Results}
\begin{figure}[pth]
\includegraphics[width=8cm]{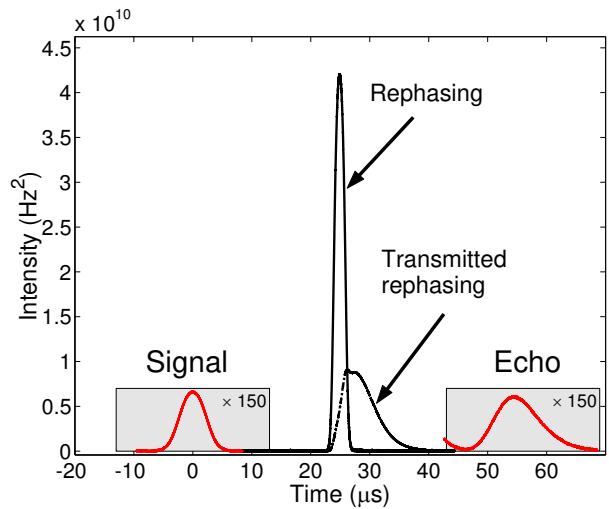}\caption{(color online) Time sequence for $1.1\pi$-rephasing pulse. The signal and the echo being very weak, we magnify their scale by a factor 150 (solid red line). The intensity is given as the square of the Rabi frequency. The signal and the rephasing pulse have a duration of $\mathrm{2.1\mu s}$ and $\mathrm{0.8\mu s}$ respectively (rms width of the gaussian). We also represent the rephasing pulse after transmission (dashed line in Arb. units).}
\label{fig:TimeSeqArticle}%
\end{figure}

We perform a 2PE experiment in the beforementioned conditions (see Fig. \ref{fig:TimeSeqArticle}). We pay special attention to an accurate calibration of the rephasing pulse area. We indeed first perform an optical nutation experiment to evaluate the exact Rabi frequency of the pulse. Comparing the signal and rephasing beam intensities, we estimate the area of the signal which is confirmed to be weak ($0.09\pi$). 

\begin{figure}[pth]
\includegraphics[width=8cm]{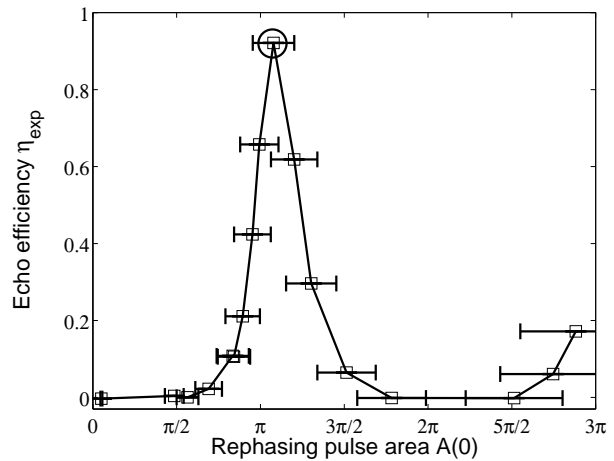}\caption{Efficiency of the echo retrieval as a function of the rephasing pulse area $A\left(0\right)$. The maximum (circle) is obtained for $A\left(0\right)=1.1\pi$ where the efficiency is 92\%, this corresponds to the situation depicted in Fig. \ref{fig:TimeSeqArticle}.}
\label{fig:EffArticle}%
\end{figure}

As expected, we observe an echo at delay $t_{12}$ after the rephasing pulse (see Fig. \ref{fig:TimeSeqArticle} ). We also carefully calibrate the efficiency. To do so, we shift the laser far from the absorption line (a few cm$^{-1}$). The corresponding intensity level represents the 100\% reference line on the measurement detector. According to the efficiency definition given by Eq. \ref{eta}, the echo is assumed to exhibit the time-reversed temporal shape of the signal. This is not exactly the case experimentally as we shall discuss later. So we define the efficiency by comparing the maxima of the two pulses:
\begin{equation}\label{etaEXP}
\eta_\mathrm{exp}\left(A\left(0\right)\right)=\left(\frac{\max_t\left(\mathcal{E}(L,t)\right)}{\max_t\left(\mathcal{S}(0,t)\right)}\right)^2
\end{equation}
By varying the incoming rephasing area, we obtain $\eta_\mathrm{exp}$ as a function of $A\left(0\right)$ (see Fig. \ref{fig:EffArticle}).

The main source of uncertainty is due to the alignement and the spatial overlap of the beams in the crystal. To quantify it and derive error bars, we estimate the typical intensity variation of the rephasing beam (waist $\mathrm{35 \mu m}$) over a length corresponding to the signal waist ($\mathrm{17 \mu m}$). This is simply given by the direct comparison of two gaussian curves (11\% in that case, which gives the horizontal error bars in Fig. \ref{fig:EffArticle}).

\subsection{Discussion}

The experimental results are qualitatively in good agreement with the expected efficiencies in Fig. \ref{fig:PlotEtaVSAi}. The main features are indeed observed. First of all the maximum efficiency is obtained for a $1.1\pi$-area, which is consistent with $\pi$, within the error bar. We also observe a reincrease of the efficiency close to $3\pi$ after a minimum at $2\pi$. We certainly predict a maximum at any odd number of $\pi$, but we are experimentally limited by the available laser power. Secondly, the curve is peaked in the vicinity of $\pi$ and cannot be fitted by a sinus-shape oscillation. This is also expected (see Fig. \ref{fig:PlotEtaVSAi}) and is due to the large optical thickness of the sample $\alpha L\simeq5$. Finally, the maximum efficiency is relatively high 0.92 for $A\left(0\right)=1.1\pi$ (circle in Fig. \ref{fig:TimeSeqArticle}). Although far below the $\eta\left(\pi\right)=146$ predicted value (Eq. \ref{etaPi}), this result demonstrates a highly efficient 2PE.

We can invoke many reasons to explain the discrepancy between the measured and the predicted efficiency values. (i) The first obvious one is the total duration of the time sequence. The echo is indeed decaying exponentially because of the coherence lifetime $T_2=50\mu s$ which has been completely neglected in our treatment. In our case $t_{12}=\mathrm{25\mu s}$ so the echo is reduced by a factor $\exp(4t_{12}/T_2)=7.4$. Without this decay, the efficiency would be much larger than unity. In section \ref{sec:QM} we shall see why $t_{12}$ is chosen to be long for this experiment. (ii) Another limitation is certainly due to the duration of the rephasing pulse. With this pulse 2.5 times shorter than the signal, the incoming pulse spectral overlap is rather good. However, propagation through the sample strongly stretches the rephasing pulse, as expected from discussion in Sec. \ref{specPI}, and as observed in Fig. \ref{fig:TimeSeqArticle}. The pulse cannot be considered as much briefer than the incoming signal all the way through the sample. An observable proof of this effect is the retrieval time of the echo (Fig. \ref{fig:TimeSeqArticle}). The retrieval should be centered on $t=2t_{12}=\mathrm{50\mu s}$. We clearly see that the echo is delayed by few microseconds. Indeed, because of stretching the rephasing pulse is no longer centered at a delay $t_{12}$ from the signal. The retrieval time is shifted accordingly. Since one of the assumption of our model is not fully verified, we then expect an efficiency reduction. (iii) In the spatial domain, the same argument is also valid. The signal is tightly focused ($\mathrm{17 \mu m})$ to ensure that its waist is smaller that the rephasing beam. The associated confocal parameter is typically two time shorter than the crystal length. So the rephasing beam does not overlap the signal uniformly all along the propagation. In other words, the rephasing area is not constant in the transverse direction. This should reduce the efficiency and broaden the peak around $\pi$ (convolution effect).

Based on this analysis, we believe our model contains all the physical ingredients to explain qualitatively the experimental results. We have given three probable explanations to interpret the quantitative discrepancy with the predicted values.

\section{Relevance for quantum memory application}
\label{sec:QM}
Our experiment has been performed in the classical domain using weak small-area pulses. It tells us however what should be the limitations in the quantum domain.

An obvious one is already present in our experiment. Since the rephasing pulse stretches while propagating through the sample, it gains a trailing tail that is not negligible as compared to the echo amplitude. As we can see in Fig. \ref{fig:TimeSeqArticle} at $t=\mathrm{45\mu s}$ the pulse tail is falling slightly before the echo comes out. That's the reason why we cannot make $t_{12}$ shorter, otherwise the echo would be submerged. With only few photons in the signal, this effect would be disastrous. As already discussed, the strong pulse distortion is not an artefact. This is a coherent propagation effect ~\cite{ruggiero}, thus a fundamental limitation. This should not be confused with the noise induced by the fluorescence, which will be another limitation at the few photons level.

The rephasing process is inherently associated with a population inversion. The decay has been neglected in our model. In practice the medium excitation will be followed by spontaneous emission. The fidelity of the 2PE as a quantum memory protocol is fundamentally limited by fluorescence. This can already be understood within the framework of the Dicke model~\cite{dicke}, i.e. without taking propagation effects into
account. Consider an ensemble of two-level systems, where
the two states are denoted $|g\rangle_k$ and $|e\rangle_k$
for the $k$-th system. Note that we are interested in the
case where the transition energy for the $g$ to $e$
transition is slightly different for different systems
(inhomogeneous broadening).

We will compare the case where the input to be stored is a
single photon to the case where there is no input (i.e.
where the input state is the vacuum). The initial state of
the atomic ensemble is $|\psi^0\rangle=|g\rangle_1
|g\rangle_2 ... |g\rangle_N$. For a vacuum input, this
state remains of course unchanged. It is then transformed
to $|\psi^N\rangle=|e\rangle_1 |e\rangle_2 ... |e\rangle_N$
by the $\pi$-pulse. For a single photon input, absorption
of the photon creates a state of the form
\begin{equation} \label{psi1}
|\psi^{1}\rangle=\frac{1}{\sqrt{N}}\left( |e\rangle_1 |g\rangle_2 ... |g\rangle_N +...+ |g\rangle_1 |g\rangle_2 ... |e\rangle_N\right)
\end{equation}
which contains a single atomic excitation. (Propagation
effects would lead to the coefficients of the $N$ terms not
being all the same.) The $\pi$-pulse transforms this state
into
\begin{equation} \label{psiminus1}
|\psi^{N-1}\rangle=\frac{1}{\sqrt{N}}\left( |g\rangle_1 |e\rangle_2 ... |e\rangle_N +...+ |e\rangle_1 |e\rangle_2 ... |g\rangle_N\right)
\end{equation}
which has $N-1$ atomic excitations.

In an inhomogeneously broadened system, the various terms
in Eqs. (\ref{psi1},\ref{psiminus1}) will acquire different
phases depending on the transition energies of the various
atoms. However, at the time of the echo all terms will be
in phase. Emission from the state $|\psi^{N-1}\rangle$
gives rise to the echo signal corresponding to a
single-photon input, whereas emission from the state
$|\psi^N\rangle$ corresponds to a vacuum input and thus
defines the noise background due to fluorescence. The
photon emission probability for a state $|\psi\rangle$ is
proportional to $||J_-|\psi\rangle||^2$, where
$J_-=\sum_{k=1}^N |g\rangle_k \langle e|_k$. This is due to
the fact that the interaction Hamiltonian between the
atomic ensemble and the relevant mode $a$ of the
electro-magnetic field (corresponding to emission in the
direction of phase matching) is proportional to
$a^{\dagger} J_-+h.c.$.

Following Ref. \cite{dicke} it is easy to see that
$||J_-|\psi^{N-1}\rangle||^2=2(N-1)$ and
$||J_-|\psi^{N}\rangle||^2=N$. As a consequence, the
probability to emit a photon at the echo time is only twice
as large for a single-photon input as for no input at all,
corresponding to a signal-to-noise ratio of one. This
severely limits the achievable fidelity of quantum state
storage.

Finally, from what we have shown is this article, we can conclude that the efficiency is actually too high for quantum memory application. We indeed observed a maximum 92\% retrieval but the efficiency can be much larger than unity with optimized conditions \cite{cornishOL00}. This amplification due to the medium inversion is precisely a propagation effect that is not considered in the beforementioned  Dicke model. This is a key ingredient to interpret our experimental results. In quantum optics terms, the statistics of the field will be modified: for one photon coming in, more than one will come out. Because the medium is inverted, it actes as a gain medium and modifies the quantum field and then again reduces the fidelity \cite{GainMed}. In that sense, the 2PE is also too efficient to be a good quantum protocol.

Those reasons are three fundamental limitations that we expect in the quantum regime. Even if the 2PE suffers from drastic drawbacks, it should be considered with attention. It is not only a historical example that helps us to understand rephasing phenomena. It has the unique ability to rephase atoms with randomly distributed level shifts, whatever the distribution structure \cite{crib3}. This a major difference as compared to CRIB (Controlled Reversible Inhomogeneous Broadening) \cite{crib1,hetet07,crib3,alexander2006} where the first initial step is a spectral tailoring of the distribution. There is not preparation of the system in the 2PE. The feat performed by the CRIB protocol is its capability to produce a rephasing of the coherence without any population inversion. As a consequence the equations of the 2PE (Eq. \ref{EfinalPI}) and the CRIB \cite{sangouard} are remarkably similar except that a minus sign accounts for the population inversion in the propagation equation. One can finally wonder if an optical manipulation would achieve a rephasing in the ground state as the CRIB does. This is \textit{a priori} not possible because even a complex optical sequence will be decomposed with rotations on the Bloch sphere, on contrary the CRIB protocol can be interpreted as a planar symmetry (detuning sign reversal). These two are then intrinsically and fundamentally different.

We have here listed the limitations of the 2PE when considered as a potential quantum storage protocol. We pointed out the pulse deformation than can be a technical issue when using strong pulse. The two other limitations are directly and fundamentally related to the medium inversion induced by the optical rephasing operation. On one side the spontaneous emission will produce a noise comparable to the retrieved signal and then deteriorate the storage fidelity. On the other side the inversion will make the medium amplifying, which mainly explains the large predicted and observed efficiency. A larger than one efficiency is also associated with a fidelity reduction for quantum fields.

\section{Conclusion}
In this article, we study the 2PE efficiency within the context of quantum storage. In this framework, we experimentally observe large efficiencies that are well explained by a simple model. Our calculations are based on a physical analysis of this specific situation. The experimental set-up has been devised to verify the underlying assumptions of the model. In 2PE, rephasing goes along with population inversion. This is a crucial ingredient of this protocol. The emitted echo is then widely amplified and can be stronger than the incoming signal. We have observed this effect. We finally conclude by analysing the potential extension of this work at low light level. We have pointed out the inherent limitations of the process.

By clarifying the physics involved in the very well-known two-pulse photon echo, we more generally tackle the problem of using strong light pulses for rephasing purposes. Our study should then be considered as a tool for the conception of new quantum storage protocols.

\end{document}